\documentclass{ws-ijmpa}
\usepackage[super,compress]{cite}
\usepackage{graphicx}
\usepackage{orcidlink}
\begin{document}
%\markboth{A. Guvendi, F. Ahmed}

\catchline{}{}{}{}{}
%
%%%%%%%%%%%%%%%%%%%%%%%%%%%%%%%%%%%%%%%%%%%%%%%%%%%%%%%%%%%%%%%%%%%%

\title{Relativistic quantum oscillator under rainbow gravity's effects in traversable wormhole with disclination}

\author{Abdullah Guvendi\orcidlink{0000-0003-0564-9899}
}

\address{Department of Basic Sciences, Erzurum Technical University, 25050, Erzurum, Turkey\\
E-mail: abdullah.guvendi@erzurum.edu.tr}

\author{Faizuddin Ahmed\orcidlink{0000-0003-2196-9622}}

\address{Department of Physics, University of Science and Technology Meghalaya, Ri-Bhoi, 793101, India\\
E-mail: faizuddinahmed15@gmail.com ; faizuddin@ustm.ac.in}

\maketitle

\begin{abstract}
In this paper, our principal objective is to investigate the impact of disclination and throat radius of a three-dimensional traversable wormhole on quantum oscillator fields.  Specifically, we focus on Perry-Mann-type wormhole with disclination while also considering the influence of rainbow gravity's. We derive the radial equation of the relativistic Klein-Gordon oscillator within this wormhole background under the effects of gravity's rainbow and the analytical eigenvalue solution is obtained using the confluent Heun function. In fact, we show that the behavior of the oscillator fields is significantly influenced not only by the presence of disclination and the throat radius but also by the parameter of rainbow gravity's. We choose various such rainbow functions to present and analyze the eigenvalue solutions of the quantum oscillator fields.

\keywords{Modified theories of gravity; Linear defects: disclination; Quantum field in curved space-time; Relativistic wave equation; Solutions of wave equations; Special functions.}
\end{abstract}

\ccode{PACS numbers: 04.50.Kd; 61.72.Lk; 04.62.+v; 03.65.Pm; 03.65.Ge; 02.30.Gp}

\section{Introduction}

Wormholes are theoretical constructs in the field of theoretical physics and general relativity. A wormhole is like a tunnel that can connect vastly different regions in space-time, potentially allowing matter or information to travel from one end to the other. These intriguing structures, also known as Einstein-Rosen bridges, were first introduced as solutions to the equations of general relativity \cite{ER}. While these solutions hinted at the possibility of such constructs existing, definitive answers regarding their stability and practicality remained elusive. The stability and feasibility of wormholes remain open questions in the realm of physics. Although they are mathematically permissible within the framework of general relativity, they pose significant challenges, including the necessity of exotic matter and the potential for instability due to quantum effects \cite{SW}. Notwithstanding their theoretical underpinnings, ongoing research in fields like quantum gravity, black hole physics, and exotic matter underscores the enduring fascination surrounding wormholes. The intricate study of how quantum particles behave in the vicinity of a wormhole in space-time is a profoundly theoretical endeavor that demands the harmonization of quantum mechanics and general relativity. Wormholes represent regions of space-time characterized by intense curvature primarily resulting from gravitational forces. Within these regions, gravitational forces can profoundly influence the dynamics of quantum particles and systems \cite{A1, A2, A3, S4}. These influences can manifest as phenomena such as time dilation and the distortion of particle trajectories \cite{TD}. In the vicinity of a wormhole, the curvature of space-time can exert substantial influence over the nature and behavior of quantum fluctuations \cite{CEW}. Effectively describing the quantum motion of particles within the extreme environment surrounding a wormhole presents a formidable theoretical challenge. Consequently, the true nature of wormholes and their intricate interactions with quantum particles and systems constitute a thriving domain of active research \cite{n1, n2, n3, n4}.

Currently, there is no complete and agreed-upon theory that successfully combines quantum mechanics and general relativity under extreme conditions. Further advancements in our understanding of quantum gravity and the development of a consistent theory that merges quantum mechanics and general relativity are necessary to gain deeper insights into this topic. Hence, as a useful way to explore the quantum motion of relativistic particles under the influence of gravitational fields, we often rely on approximations and models. Some semi-classical approaches are considered to explore the evolution of physical systems under the effects of strong gravitational fields. Doubly special relativity, presented by Amelino-Camelia \cite{AC1,AC2,AC3} and also known as rainbow gravity approach one of these approaches. This approach was generalized, by Magueijo and Smolin \cite{MS}, to doubly general relativity. Modified gravity theories or some approximations such as gravity's rainbow approximation, play a significant role in the field of theoretical physics and cosmology. These theories propose modifications to Einstein's general theory of relativity to address various challenges and questions that cannot be fully explained within the framework of general relativity. Modified gravity theories or some approximations provide a framework for testing the predictions of general relativity in various regimes. Experiments and observations that deviate from the predictions of general relativity can help constrain and refine these alternative theories, leading to a deeper understanding of gravity. Gravity's rainbow approach attempts to bridge the gap between general relativity and quantum mechanics and moreover it explores the potential effects of quantum gravity on gravitational interactions and the structure of space-time. Approximations like gravity's rainbow provide alternative frameworks for understanding gravity and addressing some of the most profound questions in cosmology and theoretical physics. While they are still subject to ongoing research and testing, they offer valuable insights and alternative explanations for various astrophysical and cosmological phenomena. Some observations demonstrate that quantum motions of ultra-energetic particles can be altered, similar to quantum fields in curved spaces. In gravity's rainbow approximation,  minimum accessible regions where the quantum gravity effects are dominant are addressed. Accordingly, in this approach, the well-known dispersion relation is modified \cite{AC2,AC3}. So the rainbow gravity approach requires using a deformed space-time background to explore the dynamics of quantum particles exposed to gravitational fields. Under this approximation, the obtained results are compatible with the usual results that can be obtained in general relativity. Rainbow gravity approach is considered for different kinds of works based on cosmology, astrophysics and relativistic quantum mechanics. Under this approximation, it was shown that temperature of black holes can be energy-dependent \cite{BH1,BH2,BH3,BH4} and moreover it was shown that rainbow gravity approach can remove information paradox of black holes \cite{IP}. Recently, quantum mechanical systems are studied in curved spaces by considering modified dispersion relation including gravity's rainbow functions. The effects of gravity's rainbow on the relativistic spin-0 bosonic oscillator system subject to Coulomb-type vector potential in topological defect-induced background \cite{kgo}, relativistic dynamics of Dirac particles in cosmic string space-time \cite{d}, Landau quantization of scalar particles in a non-trivial topology \cite{LL,LL2}, relativistic spin-1/2 oscillator \cite{KB} and Aharonov-Bohm effect \cite{KB2} were investigated besides different kind of works \cite{KS,o2,AAA,EEK,EEK2,EEK3,AFA,ABT,OM}. In this context, announced results have shown us that dynamics of relativistic quantum mechanical systems are changed when the effects of gravity's rainbow are considered.

It is also know that relativistic oscillator models such as the Dirac oscillator \cite{MM}, the KG-oscillator (KGO) \cite{SB}  are useful tools to explore the effects of curved space on the associated systems since they are exactly soluble models, in general. Such systems are a concept within the field of theoretical physics that involve the study of oscillatory behavior in systems governed by both relativistic effects and quantum mechanics. They play a significant role in understanding various aspects of particle physics and quantum field theory. The importance of relativistic oscillators lies in their ability to provide insights into several key areas of modern physics because these models help us to understand the quantization of fields, creation and annihilation of particles, and the fundamental interactions between particles and fields. The KGO provides a mathematical framework for studying the behavior of scalar bosons and scalar fields under the combined influence of relativistic effects and quantum confinement and further it results in the usual quantum oscillator in non-relativistic limit. In this context, the KGO is one of the essential tools for understanding not only the behavior of spinless particles at high energy in extreme conditions but also it acts as a bridge for the gap between classical and quantum physics in various domains. Although the KGO seems to be a fully solvable system and we have a very large literature on the dynamics of this system in curved space, we do not have a very satisfactory result for this system in a complete black hole (but see \cite{dm}) or wormhole space-time background.

In this manuscript, we investigate the influence of gravity's rainbow and parameters related to Perry-Mann-type wormhole geometry on the quantum dynamics of bosonic oscillator fields. In this study, we focus on a specific example: a static three-dimensional Perry-Mann-type wormhole embedded with disclination. We derive the radial wave equation for the Klein-Gordon oscillator within this wormhole background and subsequently obtain the analytical solution using special functions. Notably, we demonstrate that the eigenvalue solution of the oscillator field is affected by both the wormhole throat radius and the disclination parameter. Additionally, the gravity rainbow functions play a crucial role in shaping the eigenvalue, leading to a distinct energy spectrum based on chosen known rainbow functions. This manuscript is structured as follows: In {\it Section 2}, we derive and solve the relativistic Klein-Gordon oscillator in the background of Perry-Mann-type traversable wormhole with disclination. In {\it section 3}, we present the energy eigenvalues for various known rainbow functions and analyze the spectrum. Finally, {\it Section 4} contains summary and discussion of the quantum oscillator system under investigation. Throughout the analysis, we choose the system of units, where $c=1=\hbar$.

\section{Quantum oscillator field under rainbow gravity's effects in traversable wormhole background with disclination }

In this section, we center our attention on the investigation of relativistic oscillator field via the Klein-Gordon oscillator in a Perry-Mann-type wormhole background with disclination. Our specific interest lies in studying this field within the context of rainbow gravity's, while considering a wormhole structure enriched with disclination. Our primary objective is to deduce analytical solutions for the Klein-Gordon oscillator equation, and achieve this through the confluent Heun function. Following this derivation, we choose various rainbow functions and present the energy spectrum.

Therefore, we begin this section by introducing an example of Perry-Mann-type wormholes A circularly symmetric and static three-dimensional traversable wormhole was given by Perry {\it et al} \cite{PM} which is also known as Perry-Mann traversable wormhole. By choosing suitable red shift and shape functions, one of us presented an example of a three-dimensional traversable wormhole in refs. \cite{n2, n3} which we call a Perry-Mann-type wormhole with disclination given by the following line-element in the chart $(t, x, \phi)$ as \cite{PM, n2, n3} ($c=1=\hbar$)
\begin{equation}
    ds^2=-dt^2+dx^2+\mathcal{R}(x)\,d\phi^2,\quad \mathcal{R}=\alpha^2\,(x^2+a^2),
    \label{1}
\end{equation}
where $a=const$ is the wormhole throat radius and $\alpha$ relates with the angular deficit. In the presence of rainbow gravity's (RG), the above space-time (\ref{1}) can be written as 
\begin{equation}
    ds^2=-\frac{1}{f^{2}}\,dt^2+\frac{1}{h^{2}}\,\Big(dx^2+\mathcal{R}(x)\,d\phi^2\Big),
    \label{2}
\end{equation}
where $f=f(\chi)$, and $h=h(\chi)$ are the rainbow functions. For the infrared energy regimes, the rainbow functions obey the following relation $\lim_{\chi \to 0} f(\chi)=1= \lim_{\chi \to 0} h(\chi)$. Also the parameter $0\leq \chi=\frac{E}{E_{p}}\leq 1$ is the ratio of a particle's energy to Planck's energy. It is therefore convenient to define $\chi=\frac{|E|}{E_{p}}$ so that rainbow gravity's would equally affect relativistic particles and anti-particles. The covariant and contravariant form of the metric tensor for the space-time (\ref{2}) are
\begin{eqnarray}
&&g_{00}=-\frac{1}{f^{2}},\quad g_{11}=\frac{1}{h^2},\quad g_{22}=\frac{\mathcal{R}(x)}{h^2},\nonumber\\
&&g^{00}=-f^2,\quad g^{11}=h^2,\quad g^{22}=\frac{h^2}{\mathcal{R}(x)},\quad g_{ij}=0 (i\neq j)
    \label{4}
\end{eqnarray}
with it's determinant $g=-\frac{\alpha^2\,(x^2+a^2)}{f^2\,h^4}$.

The interaction of oscillator with the scalar field closely resembles the Dirac oscillator case, as described in reference \cite{MM}, employing a minimal substitution technique. In the Klein-Gordon (KG) wave equation, the oscillator is incorporated by substituting the operator $\partial_{i} \to (\partial_{i}+M\,\omega\,X_{i})$, where the vector $X_i$ takes the form $X_i=(0, x, 0)$, $\omega$ represents the oscillator frequency, and $M$ is the rest mass. Therefore, the relativistic KG-oscillator is described by the following wave equation \cite{SB,BM,EPJC,SR2,o2,o3,kgo,OM2}
\begin{eqnarray}
    \Bigg[\frac{1}{\sqrt{-g}}\,(\partial_{i}+M\,\omega\,X_{i})\Big\{(\sqrt{-g}\,g^{ij})(\partial_{j}-M\,\omega\,X_{j})\Big\}\Bigg]\,\Psi=M^2\,\Psi,
    \label{7}
\end{eqnarray}
Explicitly writing the wave equation (\ref{7}) in the space-time background (\ref{2}), we obtain the following equation
\begin{eqnarray}
    &&\Bigg[-f^2\,\frac{d^2}{dt^2}+h^2\,\Bigg\{\frac{d^2}{dx^2}+\frac{x}{x^2+a^2}\,\frac{d}{dx}-2\,M\,\omega-M^2\,\omega^2\,x^2+\frac{M\,\omega\,a^2}{x^2+a^2}\nonumber\\
    &&+\frac{1}{\alpha^2\,(x^2+a^2)}\,\frac{d^2}{d\phi^2} \Bigg\}-M^2 \Bigg]\,\Psi=0.
    \label{8}
\end{eqnarray}

In a quantum system the total wave function $\Psi (t, x, \phi)$ is expressible in terms of different variables by the method of separation of variables. Let us choose the wave function in terms of $\psi (x)$ as $\Psi (t, x, \phi)=\exp(-i\,E\,t)\,\exp(i\,\ell\,\phi)\,\psi (x)$, where $E$ is the relativistic particle's energy, and $\ell=0,\pm\,1,\pm\,2,...$ are the eigenvalues of the orbital quantum operator $-i\,\hat{\partial}_{\phi}$, and $\psi (x)$ is the radial wave function. Substituting this total wave function into the equation (\ref{8}) results the following differential equation
\begin{eqnarray}
    \psi''(x)+\frac{x}{x^2+a^2}\,\psi'(x)+\Big[\Lambda-M^2\,\omega^2\,x^2-\frac{\iota^2}{x^2+a^2}\Big]\,\psi(x)=0,
    \label{10}
\end{eqnarray}
where we set the parameters
\begin{equation}
    \Lambda=\lambda-2\,M\,\omega,\quad \lambda=\frac{1}{h^2}\,(f^2\,E^2-M^2),\quad \iota=\sqrt{\ell^2_{0}-M\,\omega\,a^2}.
    \label{11}
\end{equation}

We perform the transformation $\psi(x)=exp\Big(-\frac{1}{2}\,M\,\omega\,x^2\Big)\,H(x)$ into the Eq. (\ref{10}) results the following differential equation form
\begin{eqnarray}
    H''(x)+x\,\Big[\frac{1}{x^2+a^2}-2\,M\,\omega\Big]\,H(x)+\Big[\Pi-\frac{\tau^2}{x^2+a^2}\Big]\,H(x)=0,
    \label{13}
\end{eqnarray}
where
\begin{equation}
    \Pi=\Lambda-2\,M\,\omega=\lambda-4\,M\,\omega,\quad \tau=\sqrt{\iota^2-M\,\omega\,a^2}.
    \label{14}
\end{equation}
Changing to a new variable via $s=-\frac{x^2}{a^2}$ into the equation (\ref{13}), we obtain a second-order differential equation of the following form  
\begin{eqnarray}
    H''(s)+\Big[\zeta+\frac{\beta+1}{s}+\frac{\gamma+1}{s-1}\Big]\,H'(s)+\Big[\frac{\mu}{s}+\frac{\nu}{s-1}\Big]\,H(s)=0,
    \label{15}
\end{eqnarray}
where
\begin{eqnarray}
    \zeta=M\,\omega\,a^2,\quad \beta=-\frac{1}{2},\quad \gamma=-\frac{1}{2},\quad \mu=\frac{1}{4}\,(\tau^2-\Pi\,a^2),\quad \nu=-\frac{\tau^2}{4}.
    \label{16}
\end{eqnarray}
The differential equation (\ref{15}) is the confluent Heun equation form \cite{AR,SYS,MA} and therefore, $H(s)$ is the confluent Heun function given by
\begin{equation}
    H(s)=H_{c}\Big(M\,\omega\,a^2, -\frac{1}{2}, -\frac{1}{2}, -\frac{\Lambda\,a^2}{4}, \frac{\Lambda\,a^2}{4}-\frac{\iota^2}{4}+\frac{3}{8}; s \Big).
    \label{17}
\end{equation}

To solve the above differential equation (\ref{15}), we use a power series solution given by $H=\sum^{\infty}_{i=0}\,c_i\,s^i$ into the equation (\ref{15}). Therefore, we obtain the following recurrence relation
\begin{eqnarray}
    c_{k+2}&=&\frac{1}{2\,(k+2)(2\,k+3)}\,\Big[\Big\{4\,(k+1)(k+1-M\,\omega\,a^2)-(\iota^2-\lambda\,a^2+3\,M\,\omega\,a^2)\Big\}\,c_{k+1}\nonumber\\
    &&+\Big\{4\,M\,\omega\,a^2\,(k+1)-\lambda\,a^2\Big\}\,c_k \Big]
    \label{18}
\end{eqnarray}
with the coefficients
\begin{eqnarray}
    &&c_1=-\frac{1}{2}\,(\iota^2-\lambda\,a^2+3\,M\,\omega\,a^2)\,c_0,\nonumber\\
    &&c_2=\frac{1}{12}\Big[(\lambda\,a^2-6\,M\,\omega\,a^2-\ell^2_{0}+4)\,c_{1}+(4\,M\,\omega\,a^2\,-\lambda\,a^2)\,c_0 \Big].
    \label{19}
\end{eqnarray}

In this context, we have adopted a well-known procedure, as outlined in \cite{o2, o3, EPJC, SR2}, to derive the eigenvalue solution for the bound-state of the oscillator field. This approach has been chosen because it allows us to determine constraints on the oscillator frequency for various modes, ultimately providing us with permissible values for the eigenvalue solutions. In accordance with this method, we impose the following two conditions to conclude the power series expansion, ensuring that the resulting wave function remains regular across all regions. These conditions are
\begin{equation}
    \lambda\,a^2=4\,M\,\omega\,a^2\,(n+1)
    \label{20}
\end{equation}
And
\begin{equation}
    c_{n+1}=0\quad (n=1,2,3,....).
    \label{21}
\end{equation}

Simplification of the first condition (\ref{20}) gives us the following expression of the energy eigenvalue associated with the modes $\{n, \ell\}$ given by
\begin{equation}
    E_{n,\ell}=\pm\,\frac{1}{f(\chi)}\sqrt{M^2+4\,M\,\omega_{n,\ell}\,h^2(\chi)\,(n+1)}.
    \label{22}
\end{equation}
It is important to highlight that the energy expression provided in equation (\ref{22}) remains incomplete since a comprehensive analysis should also consider the second condition. In our examination, we focus on the lowest state of the system, characterized by $n=1$, and present the energy level and the corresponding wave function as a specific case. It's worth noting that the treatment of other states follows a similar approach. 

For the ground state defined by the mode $n=1$, from equation (\ref{22}), we obtain
\begin{equation}
    E_{1,\ell}=\pm\,\frac{1}{f}\sqrt{M^2+8\,M\,\omega_{1,\ell}\,h^2}.
    \label{23}
\end{equation}
For this same mode $n=1$, the condition (\ref{21}) implies $c_2=0$. Thus, equating the coefficient $c_2$ equals to zero, from equation (\ref{19}) we obtain
\begin{equation}
    c_1=\frac{4\,M\,\omega\,a^2}{2\,M\,\omega\,a^2-\ell^{2}_{0}+4}\,c_0.
    \label{24}
\end{equation}

Comparing $c_1$ from Eqs. (\ref{19}) and (\ref{24}), one can obtain the following expression of the oscillator frequency
\begin{equation}
    \omega_{1,\ell}=\frac{1}{3\,M\,a^2}\,\Bigg[\Big(\ell^2/\alpha^2-2\Big)\pm\sqrt{\Big(\frac{\ell^2/\alpha^2}{2}-1\Big)^2+3} \Bigg].
    \label{25}
\end{equation}

We establish a crucial constraint on the oscillator frequency, $\omega \to \omega_{1,\ell}$, which furnishes us with permissible allowed values for the ground-state energy level and its corresponding oscillator field wave function. Notably, this oscillator frequency is contingent upon both disclination $\alpha$ and the radius of the wormhole throat, $a$. Furthermore, it undergoes alterations corresponding to changes in the orbital quantum number $\ell$. In a similar vein, when dealing with higher-order modes characterized by $n \geq 2$, distinct constraints on the oscillator frequency $\omega$ emerge. These constraints offer insight into the allowable values for energy levels associated with higher states and their corresponding wave functions within the Klein-Gordon oscillator. Consequently, we have denoted the oscillator frequency in equation (\ref{22}) as $\omega \to \omega_{n,\ell}$ to account for this variation in different modes. 

Substituting this frequency $\omega_{1,\ell}$ into the equation (\ref{23}), we obtain the final expression of the bound-state energy level for the oscillator field associated with ground state given by
\begin{equation}
    f^2 (\chi)\,E^{2}_{1,\ell}-M^2=\frac{8\,h^2}{3\,a^2}\,\Bigg\{\Big(\ell^2/\alpha^2-2\Big)\pm\sqrt{\Big(\frac{\ell^2/\alpha^2}{2}-1\Big)^2+3} \Bigg\}=\frac{8\,\Delta}{3\,a^2}\,h^2 (\chi),
    \label{26}
\end{equation}
where $\Delta=\Bigg\{\Big(\ell^2/\alpha^2-2\Big)\pm\sqrt{\Big(\frac{\ell^2/\alpha^2}{2}-1\Big)^2+3} \Bigg\}$.

The corresponding ground state wave function will be
\begin{equation}
    \psi_{1,\ell}=\exp\Big(-\frac{1}{2}\,M\,\omega_{1\,\ell}\,s^2\Big)\,(c_0+c_1\,s),
    \label{27}
\end{equation}
where $\omega_{1,\ell}$ is given by (\ref{25}) and $c_1$ is given by
\begin{equation}
    c_1=\Big(\frac{\ell^2/\alpha^2}{2}-2\Big)\pm\,\sqrt{\Big(\frac{\ell^2/\alpha^2}{2}-1\Big)^2+3}.
    \label{28}
\end{equation}

Equation (\ref{26}) presents the ground-state energy level, and equation (\ref{27}) provides the corresponding wave function of the oscillator field. These solutions are derived under the constraint imposed on the oscillator frequency, as indicated in equation (\ref{25}). These calculations are conducted within the context of rainbow gravity's effects in a Perry-Mann-type wormhole background featuring disclination.

By following a similar methodology, we can determine the energy eigenvalues and corresponding wave functions for other states, specifically those characterized by modes with $n \geq 2$. It's important to note that the eigenvalue solutions of the oscillator field are subject to the influence of several factors, including the rainbow functions $f(\chi), h (\chi)$, the disclination parameter $\alpha$, and the radius of the wormhole throat, $a$. Furthermore, these eigenvalue solutions exhibit variations in response to changes in the orbital quantum number $\ell$.

\section{Energy spectrum of oscillator fields under different rainbow gravity functions}

In this section, we present an analysis of the energy spectrum of the quantum oscillator fields, which was derived in the preceding section for various rainbow functions. Our primary focus is on the pairs of gravity's rainbow functions outlined in Table 1. It is noteworthy that many of these paired functions have found applications across diverse realms of physics. Specifically, these functions have been extensively utilized to characterize the geometry of space-time within the framework of loop quantum gravity, as documented in references \cite{AC1, AC2, LL2}. Furthermore, they have played a crucial role in addressing the horizon problem, as highlighted in \cite{SHH, JM, JM2}. Notably, certain functions from this set have also arisen from the analysis of gamma-ray burst spectra at cosmological distances, as discussed in \cite{AC2}.

\begin{table}[ph]
\tbl{Various rainbow functions with $\chi=\frac{|E|}{E_p}$. $\beta_0$ is an arbitrary coefficient.}
{\begin{tabular}{ |p{1cm}|p{3cm}|p{3cm}|p{3cm}|  }
    \hline 
   Serial & Refs.   & $f(\chi)$ \quad\quad  & $h(\chi)$\\ 
    \hline\hline
 $1$ & \cite{SHH,JM,JM2}  & $\frac{1}{1-\beta_0\,\chi}$ & $\frac{1}{1-\beta_0\,\chi}$\\ 
    \hline
 $2$ & \cite{AFG} & $1$ & $1+\frac{\chi}{2}$ \\ 
    \hline
 $3$ & \cite{ZWF} & $\frac{1}{1-\beta_0\,\chi}$ & $1$ \\  
    \hline
 $4$ & \cite{AC1,AC2}  & $1$  & $\sqrt{1-\beta_0\,\chi}$ \\
 \hline
 $5$ & \cite{AC1,AC2}  & $1$   & $\sqrt{1-\beta_0\,\chi^2}$\\
 \hline
 $6$ & \cite{AC2} &  $\Big(e^{\beta_0\,\chi}-1\Big)/(\beta_0\,\chi)$ & $1$\\
 \hline
\end{tabular} \label{ta1}}
\end{table}

To unravel the behavior of spin-0 scalar particles, we will utilize the pairs of functions provided in Table 1. This endeavor involves incorporating the effects of rainbow gravity, which alters the very fabric of space-time at different energy scales, as well as considering the gravitational influences emanating from wormholes and topological defects. The relativistic energy levels of scalar particles using the functions given in the above table as follows.

\vspace{0.3cm}
{\bf Case I. Rainbow functions $f=\frac{1}{1-\beta_0\,\chi}=h$, $\chi=\frac{|E|}{E_p}$.}
\vspace{0.3cm}

Using this functions into the equation (\ref{26}), we obtain the following relation
\begin{eqnarray}
    E^{2}_{1,\ell}-M^2\,\Big(1-\frac{\beta_0}{E_p}\,|E_{1,\ell}|\Big)^2=\frac{8\,\Delta}{3\,a^2}
    \label{a1}
\end{eqnarray}
from which one can find the energy eigenvalue for $E_{1,\ell}=|E_{1,\ell}|$ as
\begin{eqnarray}
    E_{1,\ell}=\Big(1-\frac{M^2\,\beta^2_{0}}{E^{2}_p}\Big)^{-1}\Bigg[-M^2\,\frac{\beta_0}{E_p}\pm \sqrt{\frac{M^4\,\beta^2_{0}}{E^2_{p}}+\Big(M^2+\frac{8\,\Delta}{3\,a^2}\Big)\Big(1-\frac{M^2\,\beta^2_{0}}{E^{2}_p}\Big)}\Bigg].
    \label{a2}
\end{eqnarray}

\begin{figure}[b]
    \includegraphics[width=2.4in,height=1.8in]{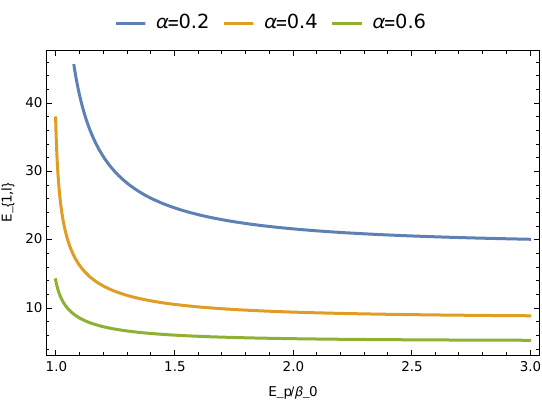}\hfill
    \includegraphics[width=2.4in,height=1.8in]{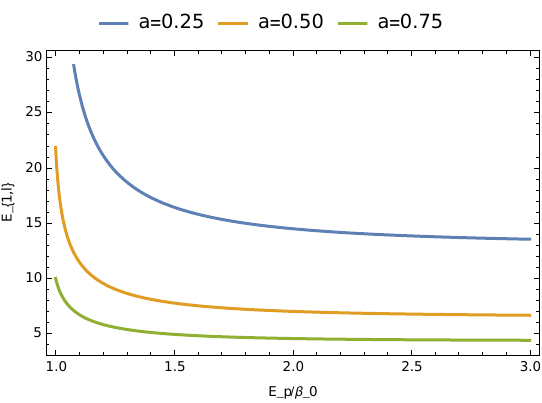}
    \caption{The energy spectrum (\ref{a2}) for $M=1=\ell$. At the left one $a=1/2$, and right one $\alpha=1/2$.}
    \label{fig: 1}
    \includegraphics[width=2.4in,height=1.8in]{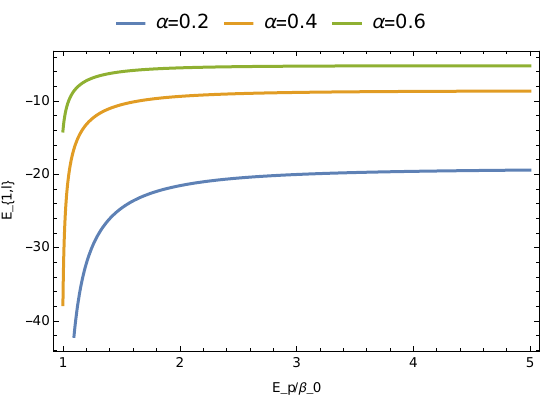}\hfill
    \includegraphics[width=2.4in,height=1.8in]{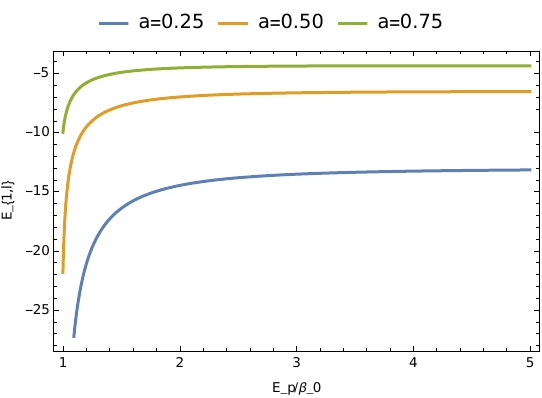}
    \caption{The energy spectrum (\ref{a3}) for $M=1=\ell$. At the left one $a=1/2$, and right one $\alpha=1/2$.}
    \label{fig: 2}
\end{figure}

And that for $E_{1,\ell}=-|E_{1,\ell}|$ will be
\begin{eqnarray}
    E_{1,\ell}=\Big(1-\frac{M^2\,\beta^2_{0}}{E^{2}_p}\Big)^{-1}\Bigg[M^2\,\frac{\beta_0}{E_p}\pm \sqrt{\frac{M^4\,\beta^2_{0}}{E^2_{p}}+\Big(M^2+\frac{8\,\Delta}{3\,a^2}\Big)\Big(1-\frac{M^2\,\beta^2_{0}}{E^{2}_p}\Big)}\Bigg].
    \label{a3}
\end{eqnarray}

We have produced Figure 1 to illustrate the behavior of the expression (\ref{a2}), and Figure 2 corresponds to (\ref{a3}). Our analysis reveals that, for a given set of values for disclination parameter $\alpha$ and the WH throat radius $a$, the energy levels exhibit a gradual decrease as the ratio $E_p/\beta_0$ increases. Importantly, this decreasing trend is more pronounced as we increase these parameters $(\alpha, a)$. Figure 2, on the other hand, shows a similar trend, but in this case, the energy levels increase as we move upwards.

\vspace{0.3cm}
{\bf Case II. Rainbow functions $f=1$, and $h=1+\frac{\beta_0}{2}\,\chi$.}
\vspace{0.3cm}

Using this functions into the equation (\ref{26}), we obtain the following relation
\begin{eqnarray}
    E^{2}_{1,\ell}-M^2=\frac{8\,\Delta}{3\,a^2}\,\Big(1-\frac{\beta_0}{E_p}\,|E_{1,\ell}|\Big)^2.
    \label{b1}
\end{eqnarray}
From above relation, one can find the energy eigenvalue 
\begin{eqnarray}
    E_{1,\ell}=\Big(1-\frac{2\,\Delta\,\beta^2_{0}}{3\,a^2\,E^2_{p}}\Big)^{-1}\Bigg[\frac{4\,\Delta\,\beta_0}{3\,a^2\,E_p}\pm \sqrt{\frac{16\,\Delta^2\,\beta^2_{0}}{9\,a^4\,E^2_{p}}+\Big(M^2+\frac{8\,\Delta}{3\,a^2}\Big)\Big(1-\frac{2\,\Delta\,\beta^2_{0}}{3\,a^2\,E^2_{p}}\Big)} \Bigg]
    \label{b2}
\end{eqnarray}
for $E_{1,\ell}=|E_{1,\ell}|$ and
\begin{eqnarray}
    E_{1,\ell}=\Big(1-\frac{2\,\Delta\,\beta^2_{0}}{3\,a^2\,E^2_{p}}\Big)^{-1}\Bigg[-\frac{4\,\Delta\,\beta_0}{3\,a^2\,E_p}\pm \sqrt{\frac{16\,\Delta^2\,\beta^2_{0}}{9\,a^4\,E^2_{p}}+\Big(M^2+\frac{8\,\Delta}{3\,a^2}\Big)\Big(1-\frac{2\,\Delta\,\beta^2_{0}}{3\,a^2\,E^2_{p}}\Big)} \Bigg]
    \label{b3}
\end{eqnarray}
for $E_{1,\ell}=-|E_{1,\ell}|$.

\begin{figure}[b]
    \includegraphics[width=2.4in,height=1.8in]{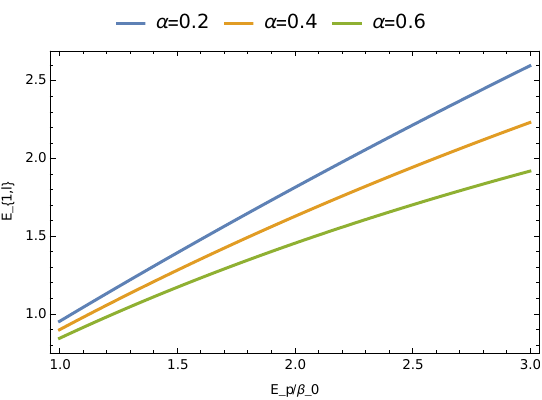}\hfill
    \includegraphics[width=2.4in,height=1.8in]{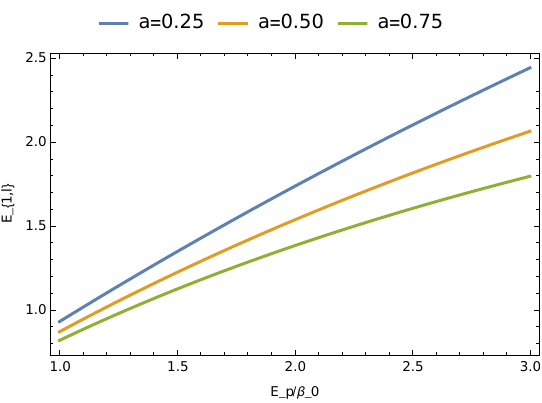}
    \caption{The energy spectrum (\ref{c2}) for $M=1=\ell$. At the left one $a=1/2$, and right one $\alpha=1/2$.}
    \label{fig: 3}
    \includegraphics[width=2.4in,height=1.8in]{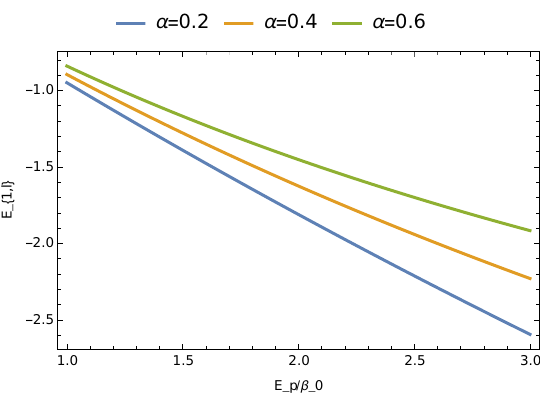}\hfill
    \includegraphics[width=2.4in,height=1.8in]{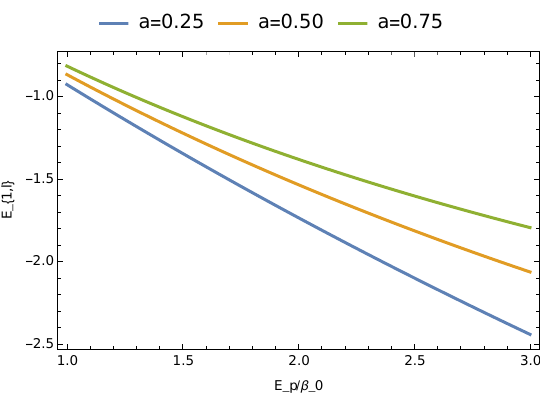}
    \caption{The energy spectrum (\ref{c3}) for $M=1=\ell$. At the left one $a=1/2$, and right one $\alpha=1/2$.}
    \label{fig: 4}
\end{figure}

\vspace{0.3cm}
{\bf Case III. Rainbow functions $f=\Big(1-\beta_0\,\chi\Big)^{-1}$, and $h=1$.}
\vspace{0.3cm}

Using this functions into the equation (\ref{26}), we obtain the following relation
\begin{eqnarray}
    \Big(1-\frac{\beta_0}{E_p}\,|E_{1,\ell}|\Big)^{-2}\,E^{2}_{1,\ell}-M^2=\frac{8\,\Delta}{3\,a^2}.
    \label{c1}
\end{eqnarray}
From above one can find the energy level 
\begin{eqnarray}
    E_{1,\ell}=\Bigg(\frac{\beta_0}{E_p}\pm\frac{1}{\sqrt{M^2+\frac{8\,\Delta}{3\,a^2}}}\Bigg)^{-1}
    \label{c2}
\end{eqnarray}
for $E_{1,\ell}=|E_{1,\ell}|$ and
\begin{eqnarray}
    E_{1,\ell}=\Bigg(-\frac{\beta_0}{E_p}\pm\frac{1}{\sqrt{M^2+\frac{8\,\Delta}{3\,a^2}}}\Bigg)^{-1}
    \label{c3}
\end{eqnarray}
that for $E_{1,\ell}=-|E_{1,\ell}|$.

We have generated Figure 3 to depict the behavior of the expression (\ref{c2}), while Figure 4 corresponds to (\ref{c3}). Our findings indicate that, for specific values of  disclination parameter $\alpha$ and the WH throat radius $a$, the energy levels exhibit an almost linear increase as the ratio $E_p/\beta_0$ grows. Notably, this upward trend becomes more pronounced as we raise the values of these parameters, namely, $(\alpha, a)$. In Figure 4, we observe a similar pattern, but in this case, the energy levels also increase, albeit in an upward direction, as we manipulate these parameters $(\alpha, a)$.

\begin{figure}[b]
    \centering
    \includegraphics[width=2.8in,height=1.6in]{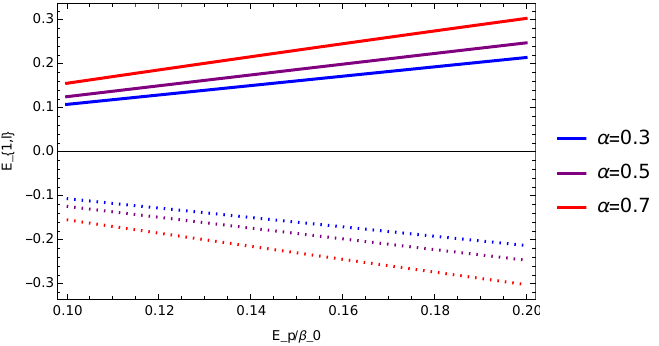}\hfill
    \includegraphics[width=2.85in,height=1.6in]{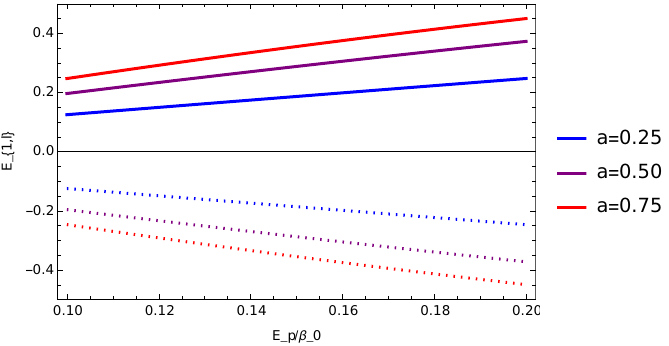}
    \caption{The energy spectrum (\ref{d2}) and (\ref{d3}) for $M=1=\ell$. At the left one $a=1$, and right one $\alpha=1/2$.}
    \label{fig: 5}
\end{figure}

\vspace{0.3cm}
{\bf Case IV. Rainbow functions $f=1$, and $h=\sqrt{1-\beta_0\,\chi}$.}
\vspace{0.3cm}

Using this functions into the equation (\ref{26}), we obtain the following relation
\begin{eqnarray}
    E^{2}_{1,\ell}-M^2=\frac{8\,\Delta}{3\,a^2}\,\Big(1-\frac{\beta_0}{E_p}\,|E_{1,\ell}|\Big).
    \label{d1}
\end{eqnarray}
Therefore, the ground-state energy eigenvalue
\begin{equation}
    E_{1,\ell}=-\frac{4\,\Delta\,\beta_0}{3\,a^2\,E_p}\pm\sqrt{\frac{16\,\Delta^2\,\beta^{2}_0}{9\,a^4\,E^{2}_p}+M^2+\frac{8\,\Delta}{3\,a^2}}
    \label{d2}
\end{equation}
for $E_{1,\ell}=|E_{1,\ell}|$ and 
\begin{equation}
    E_{1,\ell}=\frac{4\,\Delta\,\beta_0}{3\,a^2\,E_p}\pm\sqrt{\frac{16\,\Delta^2\,\beta^{2}_0}{9\,a^4\,E^{2}_p}+M^2+\frac{8\,\Delta}{3\,a^2}}
    \label{d3}
\end{equation}
that for $E_{1,\ell}=-|E_{1,\ell}|$.

We have generated Figure 5 to illustrate the behavior of both expressions (\ref{d2}) and (\ref{d3}). Our results demonstrate that, for specific values of disclination parameter $\alpha$ and the WH throat radius $a$, the energy levels display a linear increase as the ratio $E_p/\beta_0$ grows. Importantly, this upward trend becomes more prominent as we increment the values of these parameters, namely, $(\alpha, a)$.

\vspace{0.3cm}
{\bf Case V. Rainbow functions $f=1$, and $h=\sqrt{1-\beta_0\,\chi^2}$.}
\vspace{0.3cm}

Using this functions into the equation (\ref{26}), we obtain the following energy level of the oscillator field given by
\begin{equation}
    E^{2}_{1,\ell}-M^2=\frac{8\,\Delta}{3\,a^2}\,\Big(1-\frac{\beta_0}{E^{2}_p}\,E^{2}_{1,\ell}\Big)\Rightarrow E_{1,\ell}=\pm\sqrt{\frac{M^2+\frac{8\,\Delta}{3\,a^2}}{1+\frac{8\,\Delta\,\beta_0}{3\,a^2\,E^{2}_p}}}.
    \label{e1}
\end{equation}

Figure 6 has generated to depict the behavior of the expression (\ref{e1}). Our findings reveal that, for specific values of disclination parameter $\alpha$ and the WH throat radius $a$, the energy levels exhibit a gradual decrease as the ratio $E_p/\beta_0$ increases. Notably, this decreasing trend becomes more pronounced as we elevate the values of these parameters, namely, $(\alpha, a)$.

\vspace{0.3cm}
{\bf Case VI. Rainbow functions $f=\Big(e^{\beta_0\,\chi}-1\Big)/(\beta_0\,\chi)$, and $h=1$.}
\vspace{0.3cm}

Using this functions into the equation (\ref{26}), we obtain the following relation
\begin{equation}
    \frac{\Big[\exp\Big(\frac{\beta_0}{E_p}\,|E_{1,\ell}|\Big)-1\Big]^2}{\frac{\beta^{2}_0}{E^2_{p}}\,E^2_{1,\ell}}\,E^{2}_{1,\ell}-M^2=\frac{8\,\Delta}{3\,a^2}\Rightarrow E_{1,\ell}=\pm\frac{1}{\Big(\beta_0/E_p\Big)}\mbox{ln}\Bigg[1\pm \sqrt{\frac{\beta^2_{0}}{E^2_{p}}\Big(M^2+\frac{8\,\Delta}{3\,a^2}\Big)} \Bigg].
    \label{f1}
\end{equation}

We have generated Figure 7 to illustrate the behavior of the expression (\ref{f1}). Our observations indicate that, for specific values of disclination parameter $\alpha$ and the WH throat radius $a$, the energy levels demonstrate a linear increase as the ratio $E_p/\beta_0$ grows. However, it's important to note that this upward trend becomes less pronounced as we further increase the values of these parameters, specifically $(\alpha, a)$.

\section{Conclusions}

In this research, we conducted a thorough investigation of the relativistic oscillator field within the framework of rainbow gravity, considering a Perry-Mann-type wormhole background embedded with disclination. Our study revealed that, for various pairs of rainbow functions, some of which have significance in loop quantum gravity, the relativistic energy eigenvalue of the oscillator fields is influenced not only by the disclination parameter ($\alpha$) and the wormhole throat radius ($a$) but also by the rainbow parameter. In Section 2, we derived the wave equation and analytically solved the Klein-Gordon oscillator by employing the special functions. We presented the ground-state energy levels and the corresponding wave functions as a particular case for the mode associated with $n=1$ and others are in the same way. In this analysis, we have identified a constraint on the oscillator frequency $\omega$ for the mode $n=1$, which provided us the allowed values for the energy level and the corresponding wave function. Moving on to Section 3, we explored a range of rainbow functions as listed in Table 1. We then presented and analyzed the ground-state energy eigenvalues of the oscillator field for these functions. To better understand the results, we generated several graphs to visualize the trends and patterns in the data. 

Certainly, our study highlighted the sensitivity of eigenvalue solutions of the quantum oscillator field to a range of influential factors. These factors encompass the disclination, characterized by the parameter $\alpha$, as well as the WH throat radius $a$. It is evident from our findings that the rainbow parameter $\beta_0$ also exerts a significant impact on the eigenvalue solutions within the quantum system under investigation, in addition to the aforementioned parameters. Moreover, it is important to note that these eigenvalue solutions exhibit fluctuations in response to variations in the orbital quantum number $\ell$. This further emphasizes the intricate interplay of various physical parameters and quantum properties in shaping the behavior of the system under consideration.

\section*{Data Availability Statement}

No data generated or analyzed in this manuscript.

\section*{Conflicts of Interest}

Authors declare there is no conflicts of interest.

\section*{Funding statement}

No fund has received for this research.

\pagebreak

\begin{figure}[b]
    \centering
    \includegraphics[width=2.8in,height=1.6in]{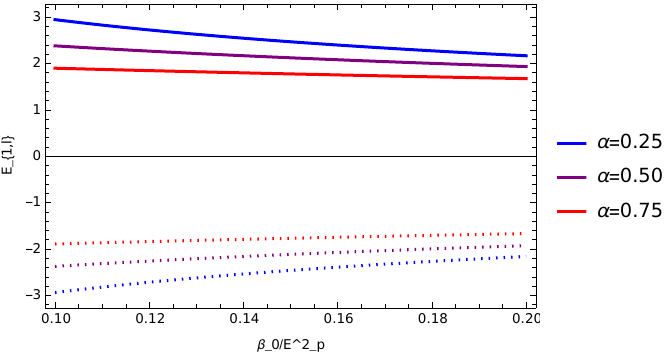}\hfill
    \includegraphics[width=3.0in,height=1.6in]{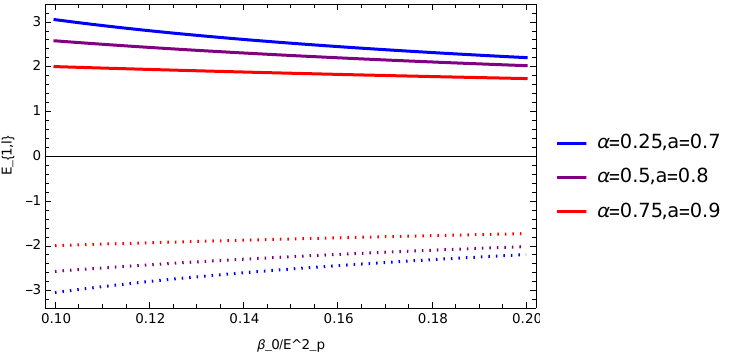}
    \caption{The energy spectrum (\ref{e1}) for $M=1=\ell$. At the left one $a=1$.}
    \label{fig: 6}
\end{figure}

\begin{figure}[b]
    \centering
    \includegraphics[width=2.9in,height=1.6in]{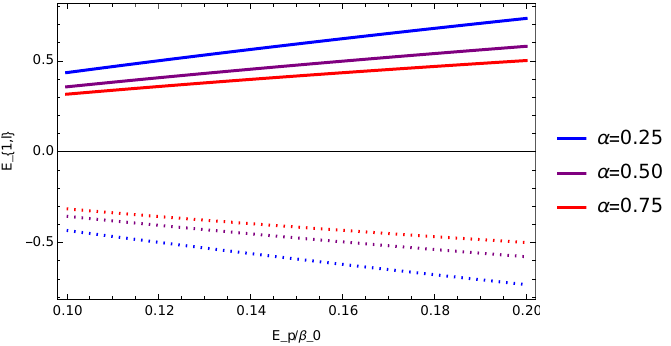}\hfill
    \includegraphics[width=2.8in,height=1.6in]{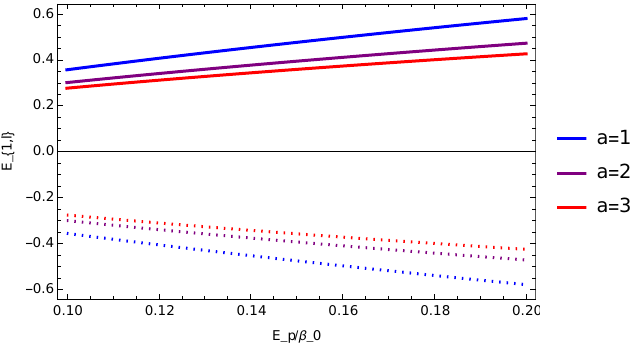}
    \caption{The energy spectrum (\ref{f1}) for $M=1=\ell$. At the left one $a=1$, and right one $\alpha=1/2$.}
    \label{fig: 7}
\end{figure}

\end{document}